\documentclass[conference]{IEEEtran}

\usepackage[font=normal]{caption}
\usepackage{subcaption}
\usepackage{enumitem}
\usepackage{amssymb}
\usepackage{tabularx} 
\IEEEoverridecommandlockouts
\usepackage{tabularray}
\usepackage{url}
\usepackage{tcolorbox}
\usepackage{listings}
\usepackage{xcolor}
\usepackage[export]{adjustbox}
\usepackage{booktabs}
\usepackage{graphicx}
\usepackage{amsmath}
\usepackage{balance}
\usepackage{fancyhdr}
\usepackage{multirow}
\usepackage{hyperref}
\hypersetup{
    colorlinks=true,
    linkcolor=blue,
    filecolor=blue,      
    urlcolor=blue,
    citecolor=blue,
    pdfpagemode=FullScreen,
    }
\usepackage{listings}

\usepackage{xcolor}
\usepackage{soul}

\definecolor{codegreen}{rgb}{0,0.6,0}
\definecolor{codegray}{rgb}{0.5,0.5,0.5}
\definecolor{codepurple}{rgb}{0.58,0,0.82}
\definecolor{backcolour}{rgb}{0.95,0.95,0.92}

\lstdefinestyle{mystyle}{
  label=code:sample,
  floatplacement=tbp,
  backgroundcolor=\color{backcolour}, commentstyle=\color{codegreen},
  keywordstyle=\color{magenta},
  numberstyle=\tiny\color{codegray},
  stringstyle=\color{codepurple},
  basicstyle=\ttfamily\footnotesize,
  breakatwhitespace=false,         
  breaklines=true,                 
  captionpos=t,                    
  keepspaces=true,                 
  numbers=left,                    
  numbersep=2pt,                  
  showspaces=false,                
  showstringspaces=false,
  showtabs=false,                  
  tabsize=1
}

\usepackage{graphicx}
\usepackage{array}
\usepackage{multirow}
\usepackage{colortbl}
\usepackage{rotating}
\AtBeginDocument{%
  \providecommand\BibTeX{{%
    \normalfont B\kern-0.5em{\scshape i\kern-0.25em b}\kern-0.8em\TeX}}}

\newcommand\change[1]{{\textcolor{blue}{#1}}}

\begin{document}

\title{VulGuard: An Unified Tool for Evaluating Just-In-Time Vulnerability Prediction Models}

\author{
    \IEEEauthorblockN{Duong Nguyen\IEEEauthorrefmark{1}, Manh Tran-Duc\IEEEauthorrefmark{1}, Thanh Le-Cong\IEEEauthorrefmark{2}, Triet Huynh Minh Le\IEEEauthorrefmark{3}, M. Ali Babar\IEEEauthorrefmark{3}, Quyet-Thang Huynh\IEEEauthorrefmark{1}}
    \IEEEauthorblockA{\IEEEauthorrefmark{1}\textit{School of Communication and Information Technology, Hanoi University of Science and Technology, Hanoi, Vietnam}
    \\ \{duong.nd215336, manh.td194616\}@sis.hust.edu.vn, thang.huynhquyet@hust.edu.vn}
    \IEEEauthorblockA{\IEEEauthorrefmark{2}\textit{School of Computing and Information Systems, The University of Melbourne, Melbourne, Australia}
    \\ congthanh.le@student.unimelb.edu.au}
    \IEEEauthorblockA{\IEEEauthorrefmark{3}\textit{School of Computer and Mathematical Sciences, The University of Adelaide, Adelaide, Australia}
    \\ \{triet.h.le, ali.babar\}@adelaide.edu.au}
}

\maketitle
\thispagestyle{fancy}

\cfoot{\thepage} 
\renewcommand{\headrulewidth}{0pt} 
\renewcommand{\footrulewidth}{0pt}
\pagestyle{fancy}
\cfoot{\thepage} 

\begin{abstract}
We present VulGuard, an automated tool designed to streamline the extraction, processing, and analysis of commits from GitHub repositories for Just-In-Time vulnerability prediction (JIT-VP) research. VulGuard automatically mines commit histories, extracts fine-grained code changes, commit messages, and software engineering metrics, and formats them for downstream analysis. In addition, it integrates several state-of-the-art vulnerability prediction models, allowing researchers to train, evaluate, and compare models with minimal setup. By supporting both repository-scale mining and model-level experimentation within a unified framework, VulGuard addresses key challenges in reproducibility and scalability in software security research. VulGuard can also be easily integrated into the CI/CD pipeline. We demonstrate the effectiveness of the tool in two influential open-source projects, FFmpeg and the Linux kernel, highlighting its potential to accelerate real-world JIT-VP research and promote standardized benchmarking. A demo video is available at: \url{https://youtu.be/j96096-pxbs}.

\end{abstract}

\section{Introduction}
\label{sec:intro}

Software vulnerabilities negatively impact the reliability, security, and functionality of software systems to an unignorable degree, leading to severe damage to both users and companies. A notable mention would be the 2024 CrowdStrike outage, a misalignment between expected field and actual input caused a cascade of system failures, affecting millions of devices and disrupting essential services worldwide \cite{crowdstrike,techtarget}. This incident highlights the significant financial and operational burdens of post-deployment detected vulnerabilities, as well as the hidden technical risks within software systems.

To mitigate these challenges, Just-In-Time Vulnerability Prediction (JIT-VP) \cite{perl2015vccfinder} has emerged as a promising approach to improving software quality assurance. At the early stages of the software development life cycle, JIT-VP techniques can identify security-threatening modifications in the software system, allowing developers to take immediate action. As a result, integrating JIT-VP into the development life cycle can improve the security inspection procedure and reduce the costs associated with future remediation.

Despite significant advancements in Just-In-Time Vulnerability Prediction (JIT-VP)~\cite{perl2015vccfinder, nguyen2024code, yang2017vuldigger}, real-world adoption remains limited. A primary obstacle lies in the complexity of data curation: extracting, cleaning, and preprocessing commits from heterogeneous and evolving software repositories is often repository-specific, error-prone, and labor-intensive. This challenge results in reduced experimental scale and inconsistent model evaluation~\cite{lomio2022just}. Furthermore, existing research rarely addresses integration with modern development workflows, thereby hindering the delivery of actionable feedback to developers and limiting the practical utility of academic models~\cite{nguyen2025toward}. To close this gap, there is a need for a unified tool that streamlines the end-to-end JIT-VP pipeline, from data collection and preprocessing to model training and evaluation, while supporting seamless integration into real-world development environments.

To address these challenges, we introduce VulGuard, a unified tool for evaluating JIT-VP techniques. This tool has been employed in the empirical study on JIT-VP presented in our recently accepted paper at ICSME 2025~\cite{nguyen2025toward}. VulGuard offers three main features: (1) dataset construction, (2) model training, and (3) model evaluation.
For dataset construction, our tool is designed to extract various features from commits, such as expert features~\cite{perl2015vccfinder,kamei2012large}, property graphs~\cite{nguyen2024code}, messages, and code changes. It also provides a tool to train and evaluate state-of-the-art JIT-VP techniques. Notably, VulGuard adopts a realistic evaluation setting that incorporates both vulnerability-related and neural commits, in accordance with the findings of our empirical study~\cite{nguyen2025toward}.
The VulGuard pipeline begins by cloning the given GitHub repositories to the local machine, then leveraging \textit{git} application to extract commit data, and the V-SZZ~\cite{bao2022v} algorithm to trace the vulnerable commits. 
 
Once the dataset is constructed, VulGuard can utilize it to train and evaluate the implemented techniques.

To summarize, key features of VulGuard include:

\begin{itemize}
    \item Construct new datasets for JIT-VP research, which are also extensible for other vulnerability analysis tasks.
    \item Support multiple programming languages, including C/C++, Java, JavaScript, and Python.
    \item Integrate multiple JIT-VP techniques to train and evaluate in real-world settings.
    \item Installable Python package with Command-line interface.
\end{itemize}

Our tool with manual is available at Github release~\cite{replication}.

\begin{figure}[t]
    \centering
    \includegraphics[width=0.9\linewidth]{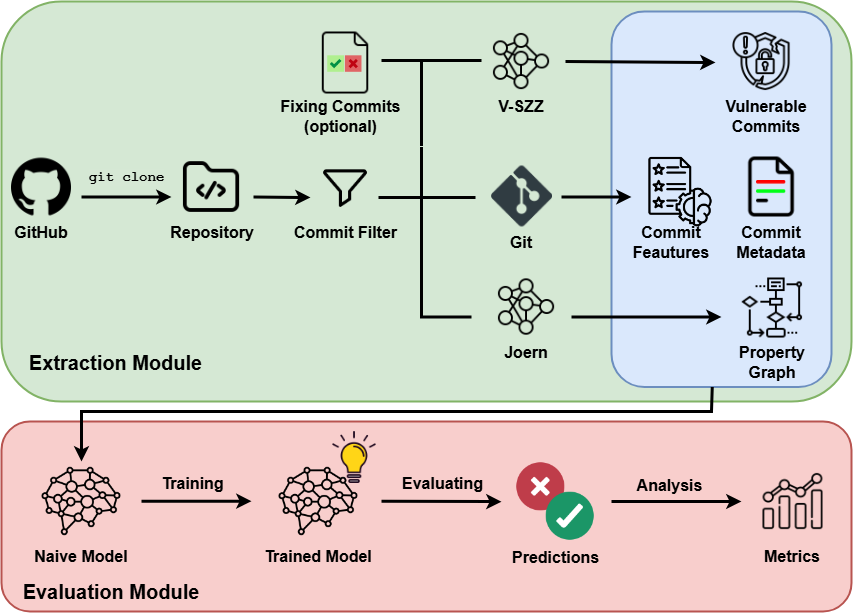}
    \caption{Architecture of VulGuard}
    \label{fig:architect}
    \vspace{-5mm}
\end{figure}

\section{Related Works}
\label{sec:background}

Zeng et al.~\cite{zeng2021deep}'s replication study is the closest to our work, which provides a codebase for extracting commit-level features and metadata. It also includes implementations of several Just-in-Time (JIT) defect prediction baselines: CC2Vec~\cite{hoang2020cc2vec}, DeepJIT~\cite{hoang2019deepjit}, DBN-JIT~\cite{yang_deep_2015}, LR-JIT~\cite{kamei2012large}, and their proposed method, LAPredict. We include some models from their work, including DeepJIT, LR-JIT, and LAPredict. However, our tool differs from their replication in multiple aspects. First, we offer user-friendly environments to facilitate adoption by both researchers and practitioners through two primary usage scenarios. (1) Our tool can be installed as a Python library, enabling integration into various software projects. (2) Besides supporting direct usage through Python library imports, we also provide an intuitive command-line interface (CLI), thereby accommodating diverse user preferences and workflows. These features provide both individual usage for research and integration usage for deployment. 
Second, our tool is designed to address the task of JIT-VP, whereas the focus of Zeng et al.'s study lies within the domain of JIT-DP, resulting in implementation-wise differences. 
Instead of employing the traditional B-SZZ algorithm~\cite{sliwerski2005changes} to identify bug-inducing commits, we utilize V-SZZ~\cite{bao2022v}, which is designed to improve the accuracy of labeling vulnerability-inducing commits.
We offer a greater variety of tools with the implementation of two more JIT-DP approaches, TLEL~\cite{yang2017tlel} and SimCom~\cite{zhou2022simple}, and three state-of-the-art JIT-VP approaches, i.e., VCCFinder \cite{perl2015vccfinder}, JITFine \cite{ni2022best}, and CodeJIT \cite{nguyen2024code}.

JITBot~\cite{khanan_jitbot:_2020} is a GitHub application for users to integrate into their own GitHub Action pipelines \cite{action}. Similarly to our work, JITBot has been created to address the problem of the lack of adoption of JIT-DP tools in CI/CD pipelines. Unfortunately, to the best of our knowledge, JITBot is no longer publicly available on GitHub. 
Moreover, while JITBot only supports the application phase of JIT-DP models using a specific built-in model, VulGuard facilitates the end-to-end development of JIT VP. This includes data mining, model training, and deployment, all within an accessible environment provided as a Python library.

\section{Architecture}
\label{sec:architecture}

VulGuard is built to streamline the data extraction process, as well as to train and evaluate prediction models for vulnerability research. VulGuard has two main modules: Extraction and Evaluation, as shown in Figure \ref{fig:architect}. We cover these modules in detail in the following subsections.

\subsection{Extraction Module}
This module of VulGuard can be divided into four main tasks: \textbf{commit collection}; \textbf{feature extraction}; \textbf{commit annotation} and \textbf{data splitting}. We also integrate a \textbf{graph builder} module, which generates a graph representation of commits.

\subsubsection{Commit Collection}
VulGuard takes input from a local Git repository. Following practices established in prior work~\cite{kim2006automatic, mcintosh2018fix,le2021deepcva}, VulGuard filters out merge commits, whitespace-only commits, and comment-only commits to focus on meaningful code modifications. In addition, only commits that involve changes in source code files based on the primary language of the repository are retained. Specifically, the tool considers files with extensions: .c/.h for C, .cpp for C++, .java for Java, .js for JavaScript, and .py for Python. 

\subsubsection{Feature Extraction}
In this task, VulGuard leverages Git to systematically collect key information from each commit, including commit messages, code changes, file-level metadata, and blame information for line-level authorship tracking. The raw data are then processed to derive relevant features that reflect code quality, developer activity, and temporal patterns~\cite{kamei2012large, perl2015vccfinder}. The features extracted in this step can be customized. All output information is serialized in .jsonl format.

\begin{table*}[t]
\centering
\caption{Regular expression used to filter patch commits provided by Zhou et al.~\cite{zhou2017automated}}
\label{table:regex}
\begin{tabular}{@{}c|p{14.5cm}@{}}
\hline
\textbf{Rule name} & \textbf{ Regular Expression} \\
\hline
strong\_vuln\_patterns & 
\texttt{(?i)(denial.of.service|\textbackslash bXXE\textbackslash b|remote.code.execution|bopen.redirect|OSVDB|\textbackslash bvuln\textbackslash b|}
\newline
\texttt{\textbackslash bCVE\textbackslash b|\textbackslash bXSS\textbackslash b|\textbackslash bReDoS\textbackslash b|\textbackslash bNVD\textbackslash b|malicious|x-frame-options|attack|cross.site|}
\newline
\texttt{exploit|directory.traversal|\textbackslash bRCE\textbackslash b|\textbackslash bdos\textbackslash b|\textbackslash bXSRF\textbackslash b|clickjack|session.fixation|}
\newline
\texttt{hijack|advisory|insecure|security|\textbackslash bcross--origin\textbackslash b|unauthori[z|s]ed|infinite.loop)} \\
\hline
medium\_vuln\_patterns & 
\texttt{(?i)(authenticat(e|ion)|brute.force|bypass|constant.time|crack|credential|\textbackslash bDoS\textbackslash b|}
\newline
\texttt{expos(e|ing)|hack|harden|injection|lockout|overflow|password|\textbackslash bPoC\textbackslash b|proof.of.concept|}
\newline
\texttt{poison|privelege|\textbackslash b(in)?secur(e|ity)|de)?serializ|spoof|timing|traversal)} \\
\hline
\end{tabular}
\end{table*}

Another key feature of VulGuard is its support for parallel execution of \texttt{git diff} and \texttt{git blame} operations. Since the metadata extraction for each commit is independent and both \texttt{git diff} and \texttt{git blame} are read-only operations, this parallelization is safe and highly effective. By leveraging concurrent processing, VulGuard significantly accelerates collecting fine-grained code changes and line-level authorship information without compromising repository integrity.

\subsubsection{Commit Annotation}

In this task, each commit is labeled as either vulnerable or non-vulnerable. Specifically, we flag the commits that have changes introducing vulnerabilities as positive, while all other commits are flagged as negative. We implement this practice to mimic the realistic conditions of the software development cycle~\cite{nguyen2025toward}.

However, accurately identifying vulnerability-inducing commits remains a challenging problem. A common practice is to identify fixing commits and then trace back to the vulnerable origins. To expedite the identification of vulnerability-fixing commits, we incorporate the regular expression proposed by Zhou et al.~\cite{zhou2017automated} (see Table \ref{table:regex}). While this technique improves efficiency, it may introduce noises. As a result, we recommend complementing \change{the tool} with a manual list of patch commits.

Next, we adopt V-SZZ~\cite{bao2022v}, an enhanced variant of the classic SZZ algorithm~\cite{sliwerski2005changes}, which traces the origin of patches to identify their corresponding inducing commits. Among the SZZ family, V-SZZ has demonstrated the highest effectiveness and has been commonly used in vulnerability analysis (e.g.,~\cite{le2024latent,cao2024snopy}). We also integrate other SZZ algorithms, such as B-SZZ~\cite{sliwerski2005changes}, AG-SZZ~\cite{kim2006automatic}, and MA-SZZ~\cite{da2016framework}, for comparative evaluations.

\subsubsection{Data Splitting}
The dataset is partitioned using a customizable ratio. By default, all commits are ordered by date to simulate continuous software development~\cite{arani2024systematic} and avoid data leakage~\cite{le2019automated} and then split using a ratio of 75/5/20\% for training/validation/testing. 

\subsubsection{Graph Builder}
Many recent JIT-VP approaches utilize structural representations for prediction. As a result, VulGuard incorporates a graph builder module that is built on the artifacts provided by CodeJIT~\cite{nguyen2024code}. In their study, Nguyen et al.~\cite{nguyen2024code} leverage Joern~\cite{joern} to generate code property graphs. This package is integrated into VulGuard with CLI. 
\subsection{Evaluation Modules}

\begin{table}[t]
\centering
\caption{Summary of approaches studied in this work and our prior work~\cite{nguyen2025toward}, and their utilized data: Expert Features (EF), Commit Messages (CM), and Commit Changes (CC).}
\label{table:model}
\begin{tabular}{l|c|ccc}
\hline
\multicolumn{1}{c|}{\multirow{2}{*}{\textbf{Models}}} &
  \multirow{2}{*}{\textbf{Technique}} &
  \multicolumn{3}{c}{\textbf{Features}} \\ \cline{3-5} 
\multicolumn{1}{c|}{} &
   &
  \multicolumn{1}{c|}{\textbf{EF}} &
  \multicolumn{1}{c|}{\textbf{CM}} &
  \textbf{CC} \\ \hline
\textbf{VCCFinder}~\cite{perl2015vccfinder} & Machine Learning  & \multicolumn{1}{c|}{\checkmark} & \multicolumn{1}{c|}{\checkmark} &            \\ \hline
\textbf{CodeJIT}~\cite{nguyen2024code} &
  Graph-based Learning &
  \multicolumn{1}{c|}{} &
  \multicolumn{1}{c|}{} &
  \checkmark \\ \hline
\textbf{LR}~\cite{kamei2012large} &
  Machine Learning &
  \multicolumn{1}{c|}{\checkmark} &
  \multicolumn{1}{c|}{} &
   \\ \hline
\textbf{TLEL}~\cite{yang2017tlel} &
  Machine Learning &
  \multicolumn{1}{c|}{\checkmark} &
  \multicolumn{1}{c|}{} &
   \\ \hline
\textbf{DeepJIT}~\cite{hoang2019deepjit} &
  Deep Learning &
  \multicolumn{1}{c|}{} &
  \multicolumn{1}{c|}{\checkmark} &
  \checkmark \\ \hline
\textbf{LAPredict}~\cite{zeng2021deep} &
  Machine Learning &
  \multicolumn{1}{c|}{\checkmark} &
  \multicolumn{1}{c|}{} &
   \\ \hline
\textbf{SimCom}~\cite{zhou2022simple}       & Ensemble Learning & \multicolumn{1}{c|}{\checkmark} & \multicolumn{1}{c|}{\checkmark} & \checkmark \\ \hline
\textbf{JITFine}~\cite{ni2022best} &
  Deep Learning &
  \multicolumn{1}{c|}{\checkmark} &
  \multicolumn{1}{c|}{\checkmark} &
  \checkmark \\ \hline
\end{tabular}
\end{table}

\begin{table}[t]
\centering
\caption{Metrics supported for model evaluation. ED is threshold dependent. ID is threshold independent.}
\label{table:metric}
\begin{tabular}{p{0.7cm}|l|p{5.5cm}}
\hline
\textbf{Type}                          & \multicolumn{1}{c|}{\textbf{Name}} & \multicolumn{1}{c}{\textbf{Description}}                                    \\ \hline
\multirow{5}{*}{DE}   & \textit{Accuracy}                  & Correct predictions out of all predictions.                   \\
                                       & \textit{Precision}                 & True positives out of all predicted positives.                 \\
                                       & \textit{Recall}                    & True positives out of all actual positives.                  \\
                                       & \textit{F1-score}                  & Harmonic mean of precision and recall,                      \\
                                       & \textit{MCC}                       & Balanced measure of prediction quality factor in class imbalance~\cite{le2024mitigating}.          \\ \hline
\multirow{2}{*}{ID} & \textit{ROC-AUC}                   & Area under ROC curve.               \\
                                       & \textit{PR-AUC}                    & Area under Precision-Recall curve.              \\ \hline
\multirow{3}{*}{Effort}                & \textit{Recall@20}                 & Percentage of actual positives found in the top 20\% of ranked predictions.  \\
                                       & \textit{Effort@20}                 & Percentage of code inspected to find top 20\% of actual positives.           \\
                                       & \textit{P-opt}                     & Measures effort saved when inspecting files in optimal versus actual order. \\ \hline
\end{tabular}
\end{table}

VulGuard provides a framework with model-level customization. Currently, we support eight prominent vulnerability prediction models: VCCFinder~\cite{perl2015vccfinder}, CodeJIT~\cite{nguyen2024code}, Logistic Regression (LR)~\cite{kamei2012large}, LAPredict~\cite{zeng2021deep}, TLEL~\cite{yang2017tlel}, DeepJIT~\cite{hoang2019deepjit}, SimCom~\cite{zhou2022simple}, and JIT-Fine~\cite{ni2022best}. These models represent a diverse set of techniques ranging from classical machine learning to deep learning and graph-based approaches. A summary of the methodology of each model is provided in Table~\ref{table:model}. Upon evaluation, VulGuard automatically computes standard classification metrics outlined in Table~\ref{table:metric}.

\begin{table*}[t]
\caption{Results of the idealized setting experiment from our empirical study~\cite{nguyen2025toward}. The highest values are in bold.}
\centering
\renewcommand{\arraystretch}{1} 
\setlength{\tabcolsep}{6pt} 
\begin{tabular}{c|c|cccccccc|c}
\hline
 & \textbf{Metric} & \textbf{VCCFinder} & \textbf{LAPredict} & \textbf{LR} & \textbf{TLEL} & \textbf{SimCom} & \textbf{DeepJIT} & \textbf{JITFine} & \textbf{CodeJIT} & \textbf{Average} \\ 
\hline
\multirow{4}{*}{\rotatebox{90}{\textbf{FFmpeg}}} 
& \text{PR-AUC}    & 0.895  & 0.558  & 0.780  & 0.850  & 0.921  & 0.906  & \textbf{0.959}  & 0.798  & 0.833  \\
& \text{MCC}       & 0.746  & 0.337  & 0.574  & 0.701  & 0.770  & 0.638  & \textbf{0.864}  & 0.579  & 0.651  \\
& \text{F1-score}  & 0.832  & 0.373  & 0.671  & 0.800  & 0.847  & 0.759  & \textbf{0.909}  & 0.716  & 0.738  \\
& \text{ROC-AUC}   & 0.948  & 0.620  & 0.865  & 0.918  & 0.954  & 0.946  & \textbf{0.980}  & 0.838  & 0.884  \\ 
\hline
\multirow{4}{*}{\rotatebox{90}{\textbf{Linux}}}  
& \text{PR-AUC}    & 0.809  & 0.610  & 0.788  & 0.829  & \textbf{0.892}  & 0.823  & 0.885  & ---   & 0.805  \\
& \text{MCC}       & 0.447  & 0.358  & 0.558  & 0.627  & 0.658  & 0.613  & \textbf{0.716}  & ---   & 0.568  \\
& \text{F1-score}  & 0.664  & 0.458  & 0.695  & 0.752  & 0.786  & 0.735  & \textbf{0.818}  & ---   & 0.701  \\
& \text{ROC-AUC}   & 0.867  & 0.699  & 0.825  & 0.881  & 0.913  & 0.873  & \textbf{0.915}  & ---   & 0.853  \\ 
\hline
\end{tabular}
\label{eval:lab}
\end{table*}

\begin{table*}[t]
\caption{Results of the realistic setting experiment from our empirical study~\cite{nguyen2025toward}. The highest values are in bold.}
\centering
\renewcommand{\arraystretch}{1} 
\setlength{\tabcolsep}{6pt} 
\begin{tabular}{c|c|cccccccc|c}
\hline
 &
  \textbf{Metric} &
  \textbf{VCCFinder} &
  \textbf{LAPredict} &
  \textbf{LR} &
  \textbf{TLEL} &
  \textbf{SimCom} &
  \textbf{DeepJIT} &
  \textbf{JITFine} &
  \textbf{CodeJIT} &
  \textbf{Average} \\ \hline
\multirow{4}{*}{\rotatebox{90}{\textbf{FFmpeg}}} & PR-AUC   & 0.071 & 0.041 & 0.093          & 0.112          & \textbf{0.134} & 0.082 & 0.111 & 0.079 & 0.091 \\
                                                 & MCC      & 0.122 & 0.067 & 0.192          & 0.169          & \textbf{0.226} & 0.138 & 0.161 & 0.135 & 0.151 \\
                                                 & F1-score & 0.132 & 0.086 & 0.176          & 0.130          & \textbf{0.231} & 0.150 & 0.156 & 0.142 & 0.150 \\
                                                 & ROC-AUC  & 0.688 & 0.591 & 0.769          & 0.795          & \textbf{0.809} & 0.746 & 0.790 & 0.721 & 0.738 \\ \hline
\multirow{4}{*}{\rotatebox{90}{\textbf{Linux}}}  & PR-AUC   & 0.013 & 0.010 & 0.023          & 0.027          & \textbf{0.031} & 0.005 & 0.005 & ---   & 0.016 \\
                                                 & MCC      & 0.030 & 0.028 & 0.070          & \textbf{0.081} & 0.073          & 0.000 & 0.000 & ---   & 0.040 \\
                                                 & F1-score & 0.036 & 0.025 & \textbf{0.039} & 0.038          & 0.034          & 0.011 & 0.011 & ---   & 0.028 \\
                                                 & ROC-AUC  & 0.588 & 0.591 & 0.746          & \textbf{0.787} & 0.779          & 0.497 & 0.497 & ---   & 0.641 \\ \hline
\end{tabular}
\label{eval:wild}
\end{table*}

\section{Usage}
\label{sec:usage}
This section covers requirements and usages of VulGuard. Example commands are available at tool release~\cite{replication}.

\subsection{Requirements}
VulGuard is designed to operate on Linux-based systems equipped with GPU acceleration. For Linux users, we support installation via the Python library and Conda environment. For other platforms, it is recommended to follow the instructions in our package to build and construct your own Docker image.

\subsection{Data Mining}
\textbf{Use case}: Extract relevent commit data for JIT-VP. In addition, automatically identify vulnerability-fixing commits, and vulnerable introducing commits.

\textbf{Preparation}: VulGuard's input of mining process is local Git repository with main languages include C/C++, Java, JavaScript, and Python. By default, Vulguard identify patch commits using regular expression (Table \ref{table:regex}). However, you can provide the tool with customize patch commits by using jsonl file with each line following this format.
\begin{lstlisting}[]
    {
        "commit_id": <commit_id>, 
        "Repository": <repo_name>
    } 
\end{lstlisting}

\textbf{Command}:
\begin{lstlisting}[language=bash]
    python -m vulguard.cli mining  \
        -dg_save_folder <save_folder> \
        -mode local \ 
        -repo_name <repository_name> \
        -repo_path <path/to/repository> \
        -repo_language <main_language_of_project> \
\end{lstlisting}

\subsection{Model Evaluation}
\textbf{Use case}: Train and evaluate implemented JIT-VP approaches. The trained models can be utilized for inference.

\textbf{Preparation}: To train and evaluate JIT-VP approaches, VulGuard leverages data extracted through a structured data mining process. By default, the dataset includes vulnerability-introducing, vulnerability-fixing, and unrelated commits. Users who wish to train or evaluate on a different dataset should provide a data file of which each line's format is:
\begin{lstlisting}[]
    {
        "commit_id": <commit_id>,
        "feature 1": <value_1>,
        ...
        "feature k": <value_k>,
        "label": <0 or 1>
    } 
\end{lstlisting}

\textbf{Command}:

\textit{Training}:
\begin{lstlisting}[language=bash]
    python -m vulguard.cli training  \
        -dg_save_folder <save_folder> \
        -model <model_name> \
        -repo_name <project_name> \
        -repo_language <main_language_of_project> \
        -epochs <epochs>
\end{lstlisting}

\textit{Testing}:
\begin{lstlisting}[language=bash]
    python -m vulguard.cli evaluating\
        -dg_save_folder <save_folder> \
        -model <model_name> \
        -repo_name <project_name> \
        -repo_language <main_language_of_project>
\end{lstlisting}

\subsection{Model Inference}
\textbf{Use case}: Utilized trained models to predict new commits. 

\textbf{Preparation}: New commits must be extracted and provided in the same format as model evaluation before inference.

\textbf{Command}:

\begin{lstlisting}[language=bash]
    python -m vulguard.cli inference  \
        -dg_save_folder <save_folder> \
        -model <model_name> \
        -repo_name <project_name> \
        -repo_language <main_language_of_project> 
\end{lstlisting}

\begin{table}[t]
\centering
\caption{Commit distribution overview for FFmpeg and the Linux kernel. The table reveals the number of vulnerability-introducing commits (\#VIC), vulnerability-fixing commits (\#VFC), vulnerability-neural commits (\#VNC), and the total number of commits in each data split.}
\label{table:stat}
\begin{tabular}{@{}llccccc@{}}
\hline
\textbf{Project} & \textbf{Split} & \textbf{\#VIC} & \textbf{\#VFC} & \textbf{\#VNC} & \textbf{\#Total} \\ \hline
\multirow{3}{*}{\textbf{FFmpeg}}       
    & Training     & 3,826 & 2,519 & 39,823  & 43,650  \\
    & Validation   & 255   & 330   & 3,242   & 3,827   \\
    & Testing      & 1,020 & 1,903 & 37,778  & 40,701  \\ \hline
\multirow{3}{*}{\textbf{Linux Kernel}} 
    & Training     & 3,461 & 1,735 & 796,965 & 800,426 \\
    & Validation   & 231   & 616   & 35,086  & 35,317  \\
    & Testing      & 922   & 1,691 & 157,039 & 157,961 \\ \hline
\textbf{Total} &               & 9,715 & 8,996 & 1,069,933 & 1,081,882 \\ \hline
\end{tabular}
\end{table}

\section{DEMONSTRATION}
\label{sec:evaluation}
VulGuard has been employed in the empirical study on JIT-VP presented
in our recently accepted paper at ICSME 2025~\cite{nguyen2025toward}. The following section summarizes our experiments and findings, and showcases VulGuard's potential application.

\subsection{Data Mining}
We apply VulGuard to mine commits from two widely used and actively maintained open-source projects: FFmpeg and the Linux kernel. These repositories are selected due to their extensive contribution histories. We collect all commits from the master branch as of September 24, 2024. The total number of commits from the two projects after the filtering process is 1,081,882. 
To accelerate the data extraction phase, we utilize parallel processing with 50 concurrent processes. It takes approximately \textbf{1 hour} for FFmpeg and roughly \textbf{12 hours} for Linux to complete data extraction. A detailed summary of the curated datasets is presented in Table~\ref{table:stat}.

\subsection{Model Evaluation}

Using VulGuard, we have conducted an empirical evaluation of the implemented JIT-VP models under two distinct settings: \texttt{ideal}, which includes only vulnerabilities and their corresponding fixing commits, and \texttt{realistic}, which also includes security-unrelated changes, resulting in data enlargement. 
The findings reveal a consistent and substantial decline in model performance when transitioning from the \texttt{ideal} to the \texttt{realistic} scenario across various evaluation metrics, i.e., PR-AUC, MCC, F1-score, and ROC-AUC. The detailed results are shown in Tables~\ref{eval:lab} and~\ref{eval:wild} for the Ideal and Realistic settings, respectively. For comprehensive analyses and discussions, please refer to our full research paper~\cite{nguyen2025toward}.

\section{Conclusion and Future Work}
\label{sec:conclusion}

We introduced VulGuard, a unified and extensible tool that automates the end-to-end process of mining, processing, and analyzing software commits for JIT-VP research. Our empirical evaluation of two influential projects, FFmpeg and Linux kernel, demonstrates the tool's practical utility and effectiveness in real-world scenarios. Looking ahead, our goal is to enhance VulGuard by incorporating ensemble learning techniques and large language models like GPT-3/4, which have been shown to work well for function-level vulnerability prediction~\cite{le2024software}, to further boost JIT-VP predictive performance as well as extend it to other vulnerability tasks~\cite{le2022survey,zhang2023survey}.


\bibliographystyle{IEEEtran}
\bibliography{main}

\begin{thebibliography}{10}
\providecommand{\url}[1]{#1}
\csname url@samestyle\endcsname
\providecommand{\newblock}{\relax}
\providecommand{\bibinfo}[2]{#2}
\providecommand{\BIBentrySTDinterwordspacing}{\spaceskip=0pt\relax}
\providecommand{\BIBentryALTinterwordstretchfactor}{4}
\providecommand{\BIBentryALTinterwordspacing}{\spaceskip=\fontdimen2\font plus
\BIBentryALTinterwordstretchfactor\fontdimen3\font minus \fontdimen4\font\relax}
\providecommand{\BIBforeignlanguage}[2]{{%
\expandafter\ifx\csname l@#1\endcsname\relax
\typeout{** WARNING: IEEEtran.bst: No hyphenation pattern has been}%
\typeout{** loaded for the language `#1'. Using the pattern for}%
\typeout{** the default language instead.}%
\else
\language=\csname l@#1\endcsname
\fi
#2}}
\providecommand{\BIBdecl}{\relax}
\BIBdecl

\bibitem{crowdstrike}
\BIBentryALTinterwordspacing
CrowdStrike, ``Crowdstrike outage report,'' 2024. [Online]. Available: \url{https://www.crowdstrike.com/wp-content/uploads/2024/08/Channel-File-291-Incident-Root-Cause-Analysis-08.06.2024.pdf}
\BIBentrySTDinterwordspacing

\bibitem{techtarget}
\BIBentryALTinterwordspacing
TechTarget, ``Crowdstrike outage damage,'' 2024. [Online]. Available: \url{https://www.techtarget.com/whatis/feature/Explaining-the-largest-IT-outage-in-history-and-whats-next}
\BIBentrySTDinterwordspacing

\bibitem{perl2015vccfinder}
H.~Perl, S.~Dechand, M.~Smith, D.~Arp, F.~Yamaguchi, K.~Rieck, S.~Fahl, and Y.~Acar, ``Vccfinder: Finding potential vulnerabilities in open-source projects to assist code audits,'' in \emph{the 22nd ACM SIGSAC conference on computer and communications security}, 2015, pp. 426--437.

\bibitem{nguyen2024code}
S.~Nguyen, T.-T. Nguyen, T.~T. Vu, T.-D. Do, K.-T. Ngo, and H.~D. Vo, ``Code-centric learning-based just-in-time vulnerability detection,'' \emph{Journal of Systems and Software}, vol. 214, p. 112014, 2024.

\bibitem{yang2017vuldigger}
L.~Yang, X.~Li, and Y.~Yu, ``Vuldigger: A just-in-time and cost-aware tool for digging vulnerability-contributing changes,'' in \emph{GLOBECOM 2017-2017 IEEE Global Communications Conference}.\hskip 1em plus 0.5em minus 0.4em\relax IEEE, 2017, pp. 1--7.

\bibitem{lomio2022just}
F.~Lomio, E.~Iannone, A.~De~Lucia, F.~Palomba, and V.~Lenarduzzi, ``Just-in-time software vulnerability detection: Are we there yet?'' \emph{Journal of Systems and Software}, vol. 188, p. 111283, 2022.

\bibitem{nguyen2025toward}
D.~Nguyen, T.~Le-Cong, T.~Huynh Minh~Le, M.~A. Babar, and Q.-T. Huynh, ``Toward realistic evaluations of just-in-time vulnerability prediction,'' in \emph{the 41st International Conference on Software Maintenance and Evolution}.\hskip 1em plus 0.5em minus 0.4em\relax IEEE, 2025.

\bibitem{kamei2012large}
Y.~Kamei, E.~Shihab, B.~Adams, A.~E. Hassan, A.~Mockus, A.~Sinha, and N.~Ubayashi, ``A large-scale empirical study of just-in-time quality assurance,'' \emph{IEEE Transactions on Software Engineering}, vol.~39, no.~6, pp. 757--773, 2012.

\bibitem{bao2022v}
L.~Bao, X.~Xia, A.~E. Hassan, and X.~Yang, ``V-szz: automatic identification of version ranges affected by cve vulnerabilities,'' in \emph{the 44th International Conference on Software Engineering}, 2022, pp. 2352--2364.

\bibitem{replication}
\BIBentryALTinterwordspacing
D.~Nguyen, M.~Tran-Duc, T.~Le-Cong, T.~Huynh Minh~Le, M.~A. Babar, and Q.-T. Huynh, ``Vulguard: An unified framework for evaluating just-in-time vulnerability prediction models,'' 2025. [Online]. Available: \url{https://github.com/AI4Code-HUST/VulGuard}
\BIBentrySTDinterwordspacing

\bibitem{zeng2021deep}
Z.~Zeng, Y.~Zhang, H.~Zhang, and L.~Zhang, ``Deep just-in-time defect prediction: how far are we?'' in \emph{the 30th ACM SIGSOFT International Symposium on Software Testing and Analysis}, 2021, pp. 427--438.

\bibitem{hoang2020cc2vec}
T.~Hoang, H.~J. Kang, D.~Lo, and J.~Lawall, ``Cc2vec: Distributed representations of code changes,'' in \emph{the ACM/IEEE 42nd international conference on software engineering}, 2020, pp. 518--529.

\bibitem{hoang2019deepjit}
T.~Hoang, H.~K. Dam, Y.~Kamei, D.~Lo, and N.~Ubayashi, ``Deepjit: an end-to-end deep learning framework for just-in-time defect prediction,'' in \emph{2019 IEEE/ACM 16th International Conference on Mining Software Repositories (MSR)}.\hskip 1em plus 0.5em minus 0.4em\relax IEEE, 2019, pp. 34--45.

\bibitem{yang_deep_2015}
\BIBentryALTinterwordspacing
X.~Yang, D.~Lo, X.~Xia, Y.~Zhang, and J.~Sun, ``Deep {Learning} for {Just}-in-{Time} {Defect} {Prediction},'' in \emph{2015 {IEEE} {International} {Conference} on {Software} {Quality}, {Reliability} and {Security}}.\hskip 1em plus 0.5em minus 0.4em\relax Vancouver, BC, Canada: IEEE, Aug. 2015, pp. 17--26. [Online]. Available: \url{http://ieeexplore.ieee.org/document/7272910/}
\BIBentrySTDinterwordspacing

\bibitem{sliwerski2005changes}
J.~{\'S}liwerski, T.~Zimmermann, and A.~Zeller, ``When do changes induce fixes?'' \emph{ACM sigsoft software engineering notes}, vol.~30, no.~4, pp. 1--5, 2005.

\bibitem{yang2017tlel}
X.~Yang, D.~Lo, X.~Xia, and J.~Sun, ``Tlel: A two-layer ensemble learning approach for just-in-time defect prediction,'' \emph{Information and Software Technology}, vol.~87, pp. 206--220, 2017.

\bibitem{zhou2022simple}
X.~Zhou, D.~Han, and D.~Lo, ``Simple or complex? together for a more accurate just-in-time defect predictor,'' in \emph{the 30th IEEE/ACM International Conference on Program Comprehension}, 2022, pp. 229--240.

\bibitem{ni2022best}
C.~Ni, W.~Wang, K.~Yang, X.~Xia, K.~Liu, and D.~Lo, ``The best of both worlds: integrating semantic features with expert features for defect prediction and localization,'' in \emph{the 30th ACM Joint European Software Engineering Conference and Symposium on the Foundations of Software Engineering}, 2022, pp. 672--683.

\bibitem{khanan_jitbot:_2020}
\BIBentryALTinterwordspacing
C.~Khanan, W.~Luewichana, K.~Pruktharathikoon, J.~Jiarpakdee, C.~Tantithamthavorn, M.~Choetkiertikul, C.~Ragkhitwetsagul, and T.~Sunetnanta, ``\BIBforeignlanguage{en}{{JITBot}: an explainable just-in-time defect prediction bot},'' in \emph{\BIBforeignlanguage{en}{the 35th {IEEE}/{ACM} {International} {Conference} on {Automated} {Software} {Engineering}}}.\hskip 1em plus 0.5em minus 0.4em\relax Virtual Event Australia: ACM, Dec. 2020, pp. 1336--1339. [Online]. Available: \url{https://dl.acm.org/doi/10.1145/3324884.3415295}
\BIBentrySTDinterwordspacing

\bibitem{action}
\BIBentryALTinterwordspacing
{GitHub}, ``Github action.'' [Online]. Available: \url{https://github.com/features/actions}
\BIBentrySTDinterwordspacing

\bibitem{kim2006automatic}
S.~Kim, T.~Zimmermann, K.~Pan, E.~James~Jr \emph{et~al.}, ``Automatic identification of bug-introducing changes,'' in \emph{the 21st IEEE/ACM international conference on automated software engineering (ASE'06)}.\hskip 1em plus 0.5em minus 0.4em\relax IEEE, 2006, pp. 81--90.

\bibitem{mcintosh2018fix}
S.~McIntosh and Y.~Kamei, ``Are fix-inducing changes a moving target? a longitudinal case study of just-in-time defect prediction,'' in \emph{the 40th international conference on software engineering}, 2018, pp. 560--560.

\bibitem{le2021deepcva}
T.~H.~M. Le, D.~Hin, R.~Croft, and M.~A. Babar, ``Deepcva: Automated commit-level vulnerability assessment with deep multi-task learning,'' in \emph{2021 36th IEEE/ACM International Conference on Automated Software Engineering (ASE)}.\hskip 1em plus 0.5em minus 0.4em\relax IEEE, 2021, pp. 717--729.

\bibitem{zhou2017automated}
Y.~Zhou and A.~Sharma, ``Automated identification of security issues from commit messages and bug reports,'' in \emph{the 11th joint meeting on foundations of software engineering}, 2017, pp. 914--919.

\bibitem{le2024latent}
T.~H.~M. Le, X.~Du, and M.~A. Babar, ``Are latent vulnerabilities hidden gems for software vulnerability prediction? an empirical study,'' in \emph{the 21st International Conference on Mining Software Repositories}, 2024, pp. 716--727.

\bibitem{cao2024snopy}
S.~Cao, X.~Sun, X.~Wu, D.~Lo, L.~Bo, B.~Li, X.~Liu, X.~Lin, and W.~Liu, ``Snopy: Bridging sample denoising with causal graph learning for effective vulnerability detection,'' in \emph{the 39th IEEE/ACM International Conference on Automated Software Engineering}, 2024, pp. 606--618.

\bibitem{da2016framework}
D.~A. Da~Costa, S.~McIntosh, W.~Shang, U.~Kulesza, R.~Coelho, and A.~E. Hassan, ``A framework for evaluating the results of the szz approach for identifying bug-introducing changes,'' \emph{IEEE Transactions on Software Engineering}, vol.~43, no.~7, pp. 641--657, 2016.

\bibitem{arani2024systematic}
A.~K. Arani, T.~H.~M. Le, M.~Zahedi, and M.~A. Babar, ``Systematic literature review on application of learning-based approaches in continuous integration,'' \emph{IEEE Access}, 2024.

\bibitem{le2019automated}
T.~H.~M. Le, B.~Sabir, and M.~A. Babar, ``Automated software vulnerability assessment with concept drift,'' in \emph{the 16th International Conference on Mining Software Repositories (MSR)}.\hskip 1em plus 0.5em minus 0.4em\relax IEEE, 2019, pp. 371--382.

\bibitem{joern}
\BIBentryALTinterwordspacing
Joernio, ``Joern.'' [Online]. Available: \url{https://github.com/joernio/joern}
\BIBentrySTDinterwordspacing

\bibitem{le2024mitigating}
T.~H.~M. Le and M.~Ali~Babar, ``Mitigating data imbalance for software vulnerability assessment: Does data augmentation help?'' in \emph{the 18th ACM/IEEE International Symposium on Empirical Software Engineering and Measurement}, 2024, pp. 119--130.

\bibitem{le2024software}
T.~H.~M. Le, M.~A. Babar, and T.~H. Thai, ``Software vulnerability prediction in low-resource languages: An empirical study of codebert and chatgpt,'' in \emph{the 28th International Conference on Evaluation and Assessment in Software Engineering}, 2024, pp. 679--685.

\bibitem{le2022survey}
T.~H.~M. Le, H.~Chen, and M.~A. Babar, ``A survey on data-driven software vulnerability assessment and prioritization,'' \emph{ACM Computing Surveys}, vol.~55, no.~5, pp. 1--39, 2022.

\bibitem{zhang2023survey}
Q.~Zhang, C.~Fang, Y.~Ma, W.~Sun, and Z.~Chen, ``A survey of learning-based automated program repair,'' \emph{ACM Transactions on Software Engineering and Methodology}, vol.~33, no.~2, pp. 1--69, 2023.

\end{thebibliography}

\end{document}